\documentclass{desyproc}

\begin{document}
\title{LaTeX template for "Patras 2012" proceedings}

\title{\vspace{-2.05cm}
\hfill{\small{DESY 12-217}}\\[1.27cm]
Update on Hidden Sectors with Dark Forces and Dark Matter}

\author{{\slshape Sarah Andreas}\\[1ex]
Deutsches Elektronen-Synchrotron DESY, Notkestrasse 85, 22607 Hamburg, Germany}

\contribID{andreas\_sarah}

\desyproc{DESY-PROC-2012-04}
\acronym{Patras 2012} 
\doi  

\maketitle

\begin{abstract}
Recently there has been much interest in hidden sectors, especially in the context of dark matter and ``dark forces'', since they are a common feature of beyond standard model scenarios like string theory and SUSY and additionally exhibit interesting phenomenological aspects. Various laboratory experiments place limits on the so-called hidden photon and  continuously further probe and constrain the parameter space; an updated overview is presented here. Furthermore, for several hidden sector models with light dark matter we study the viability with respect to the relic abundance and direct detection experiments.
\end{abstract}

\section{Introduction}

Hidden sectors, especially with light extra U(1) gauge bosons, arise naturally in different standard model extensions like string theory and supersymmetry. They recently attracted much attention since the so-called ``hidden photon'' could account for the discrepancy in the muon anomalous magnetic moment~\cite{Pospelov:2008zw} and, in addition, dark matter (DM) interacting through a light dark force has been considered as explanation for recent astrophysical observations~\cite{Fayet:2007ua,Pospelov:2007mp,ArkaniHamed:2008qn,Cheung:2009qd}.

The only interaction between the hidden sector and the standard model is through kinetic mixing of the hidden photon $\gamma'$ with the ordinary photon, generated for example by integrating out heavy particles charged under both U(1)s~\cite{Holdom:1985ag,Foot:1991bp}. Therefore we assume the kinetic mixing, parametrized by $\chi$, to be of the order of a loop factor $\chi \sim 10^{-3}-10^{-4}$ and impose the following relation with the hidden gauge coupling $g_h$,
\begin{equation}
\chi = \frac{g_Y g_h}{16 \pi^2} \ \kappa ,\label{eq-kappa}
\end{equation}
where $\kappa$ is an $\mathcal{O}(1)$ factor depending on the masses of the particles in the loop.

The low energy effective Lagrangian of the most simple hidden sector containing only an
extra U(1) symmetry and the corresponding hidden photon $\gamma'$ is then given by
\begin{equation}
{\cal L} = -\frac{1}{4}F_{\mu \nu}F^{\mu \nu} - \frac{1}{4}X_{\mu \nu}X^{\mu \nu}
- \frac{\chi}{2} X_{\mu \nu} F^{\mu \nu} + \frac{m_{\gamma^{\prime}}^2}{2}  X_{\mu}X^\mu + g_Y j^\mu_{\mathrm{em}} A_\mu,
\end{equation}
with the field strength $F_{\mu\nu}$ of the ordinary electromagnetic field $A_\mu$ and the field strength $X_{\mu\nu}$ of the hidden U(1) field $X_\mu$. The mass $m_{\gamma'}$ of the hidden photon, generated either by the Higgs or the St\"{u}ckelberg mechanism, can naturally be obtained at the GeV-scale~\cite{Baumgart:2009tn,Goodsell:2009xc,Cicoli:2011yh,Andreas:2011xf,Andreas:2011in}. 

In this paper we give an update on the experimental constraints on $\chi$ vs $m_{\gamma'}$ in Sec.~\ref{sec-HP} and discuss in Sec.~\ref{sec-DM} different models for DM interacting through a hidden photon with respect to the relic density and signature in direct detection, based on our analysis presented in~\cite{Andreas:2011in}.

\section{Currents status of limits on hidden photons}\label{sec-HP}

\begin{wrapfigure}[15]{r}{0.69\textwidth}
\vspace{-0.35cm} \centerline{
\includegraphics[height=0.33\textwidth]{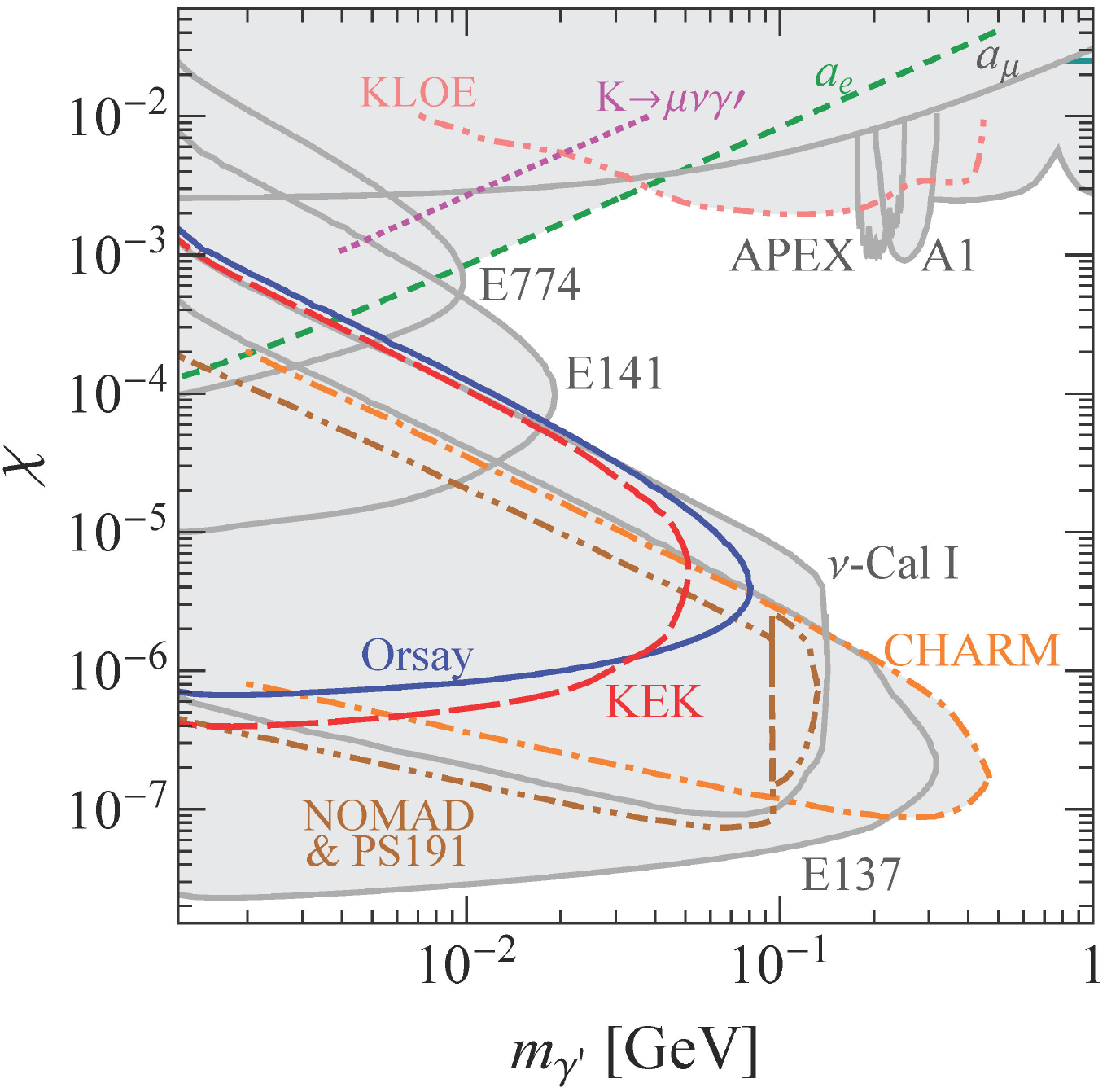}
\includegraphics[height=0.33\textwidth]{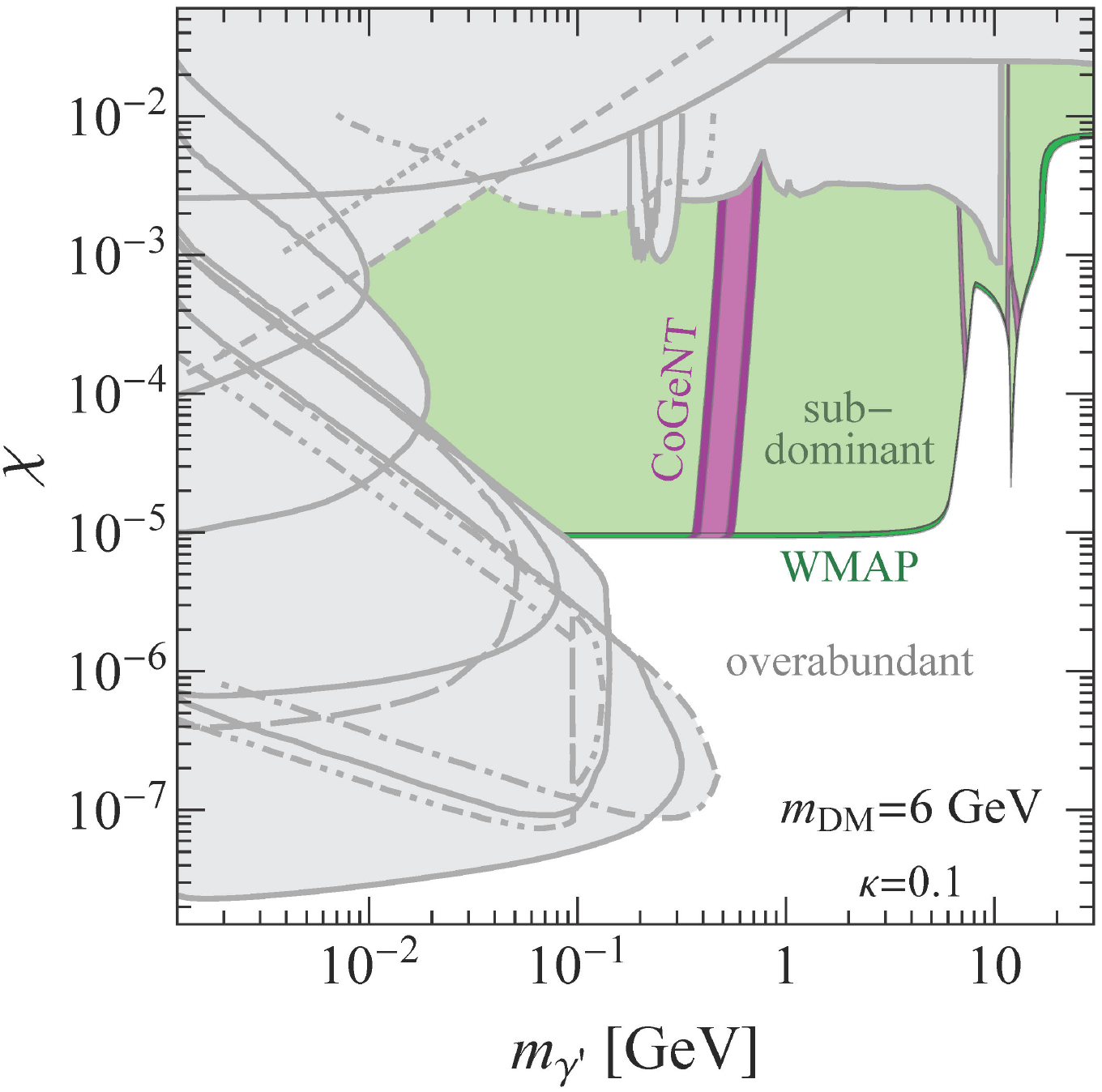}}
\vspace{-0.4cm} \caption{\textit{left:} Limits on hidden photons, recent ones in colour. \textit{right:} DM relic abundance (green) and CoGeNT region (purple) for a 6 GeV Dirac fermion DM in the toy-model with $\kappa=0.1$.} \label{fig-HPlimits}
\vspace{-0.35cm}
\end{wrapfigure}

Different laboratory experiments constrain the hidden photon through its coupling to SM fermions. In addition to the limits summarized in~\cite{Andreas:2011xf} several new exclusions arose over the last year. They will be described in the following and are shown in colour in Fig.~\ref{fig-HPlimits} on the left.

Neutrino experiments at CERN place constraints on hidden photons since neutral pseudoscalar mesons generated in the proton beams can produce spin-1 particles in their radiative decay. Limits have been derived for NOMAD and PS191 using $\pi^0$ decays in~\cite{Gninenko:2011uv} (brown double-dash-dotted line) and for CHARM from $\eta$ and $\eta'$ decays in~\cite{Gninenko:2012eq} (orange dash-dotted line). The constraint from the anomalous magnetic moment of the electron $a_e$, which was derived in~\cite{Pospelov:2008zw} together with the one of the muon, has recently been updated in~\cite{Davoudiasl:2012ig,Endo:2012hp} (green dashed line). Another limit, derived in~\cite{Beranek:2012ey} from the Kaon decay $K\rightarrow\mu\nu\gamma'$ (magenta dotted line), would have improved the former constraint of $a_e$ but is not competitive with the improved one.

Constraints from electron beam dump experiments searching for the decay of hidden photons produced in Bremsstrahlung have been derived in~\cite{Bjorken:2009mm} for the E141 and E137 experiments at SLAC as well as E774 at Fermilab. In~\cite{Andreas:2012mt} we considered two additional experiments at KEK in Japan~\cite{Konaka:1986cb} and at the Orsay Linac in France~\cite{Davier:1989wz}. The former was originally searching for axionlike particles and for the latter, designed to look for light Higgs bosons, the possibility to constrain $U$-bosons (similar to $\gamma'$ but additional axial couplings) had also been suggested in~\cite{Bouchiat:2004sp}. In addition to the new limits we reanalysed the earlier ones including the experimental acceptances which we obtained from Monte Carlo simulations with \textsc{MadGraph}. The resulting renewed 
limits are shown in gray in Fig.~\ref{fig-HPlimits} (left) together with the new ones from KEK (red long-dashed line) and Orsay (blue solid line).

The \mbox{KLOE-2} experiment at the DA$\phi$NE $\phi$-factory in Frascati has recently released a new limit~\cite{Babusci:2012cr} (pink dash-double-dotted line) for hidden photons produced in $\phi$ decays and subsequently decaying into $e^+e^- $ which improved their previous bound~\cite{Archilli:2011zc}.

\section{Hidden sector with hidden photon and dark matter}\label{sec-DM}

\subsubsection*{Toy-model: Dirac fermion as dark matter candidate}%
We assume the simplest possible hidden sector containing a hidden photon and one additional Dirac fermion as dark matter candidate (cf.~\cite{Pospelov:2007mp,Andreas:2011in,Chun:2010ve,Mambrini:2010dq,Hooper:2012cw}). The hidden photon mediates both the DM annihilation and the DM scattering on nuclei thereby determining both the DM relic abundance and direct detection rate. Applying relation~(\ref{eq-kappa}) with $\kappa=0.1$ we find that DM with a mass of 6 GeV has the correct relic abundance in the dark green band in Fig.~\ref{fig-HPlimits} (right) while it provides a subdominant contribution to the total DM density in the light green area. In the latter case, the spin-independent (SI) scattering cross section $\sigma_\mathrm{SI}$ was rescaled by the relic abundance. We find that the cross sections in the purple band (90\% CL lighter, 99\% CL darker purple) can explain the CoGeNT signal for a Standard Halo Model. Constraints from CDMS 
\begin{wrapfigure}[10]{r}{0.34\textwidth}
\vspace{-0.45cm} 
\includegraphics[width=0.33\textwidth]{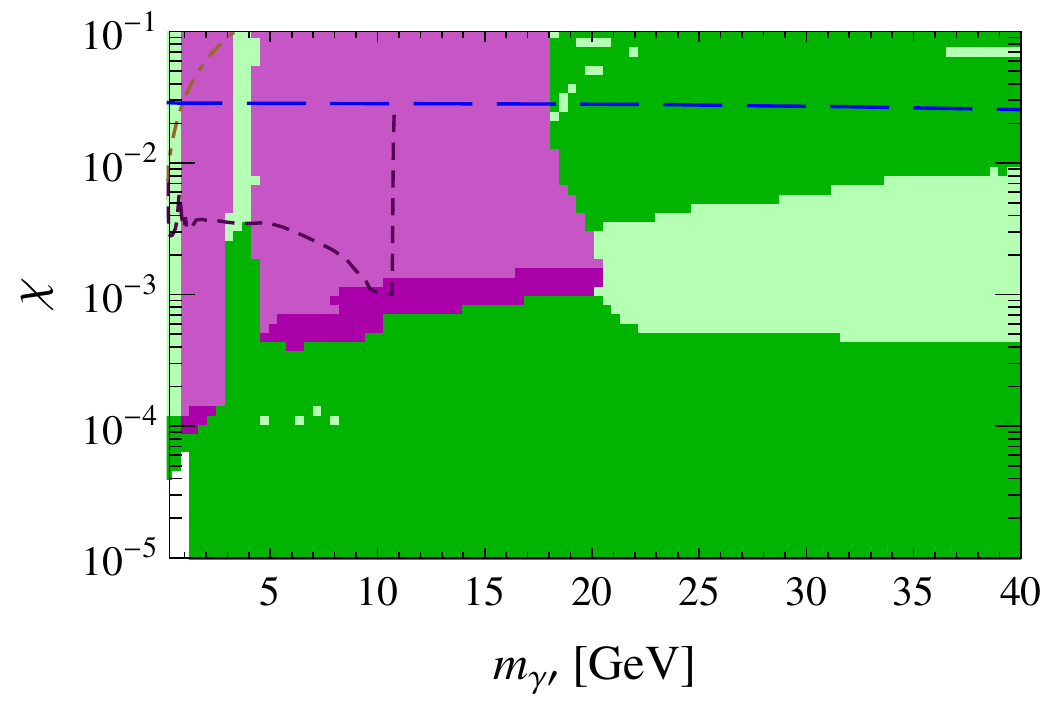} \vspace{-0.4cm} 
\caption{Scatter plot for the toy-model with a Dirac fermion.} \label{fig-ScattPlot}
\vspace{-0.35cm}
\end{wrapfigure}
and XENON do not apply to DM masses as light as 6 GeV and are thus not displayed. The excluded grey areas are the same as displayed in the left plot. The DM mass, being a free parameter of the model, can be scanned over and thereby allows to cover the entire parameter space as shown in Fig.~\ref{fig-ScattPlot}. Again, dark green corresponds to the correct relic abundance, light green to subdominant DM and purple to points in agreement with CoGeNT (all points are allowed by all other direct detection limits). Further results for different DM masses, values of $\kappa$ and halo models are presented in~\cite{Andreas:2011in}.

\subsubsection*{Supersymmetric model: Majorana and Dirac fermion as dark matter candidates}
With the aim to construct more sophisticated and better motivated models, we consider the simplest possible anomaly-free supersymmetric hidden sector model without adding dimensionful supersymmetric quantities. Therefore, we take the superpotential $W \supset \lambda_S S H_+ H_-$ with the dimensionless coupling~$\lambda_S$ and three chiral superfields $S, H_+, H_-$ where $H_+$ and $H_-$ are charged under the hidden U(1). We assume the MSSM in the visible sector, however, the DM phenomenology of the hidden sector is largely independent of this choice and instead determined by 
\begin{wrapfigure}{r}{0.69\textwidth}
\vspace{-0.25cm} \centerline{
\includegraphics[width=0.33\textwidth]{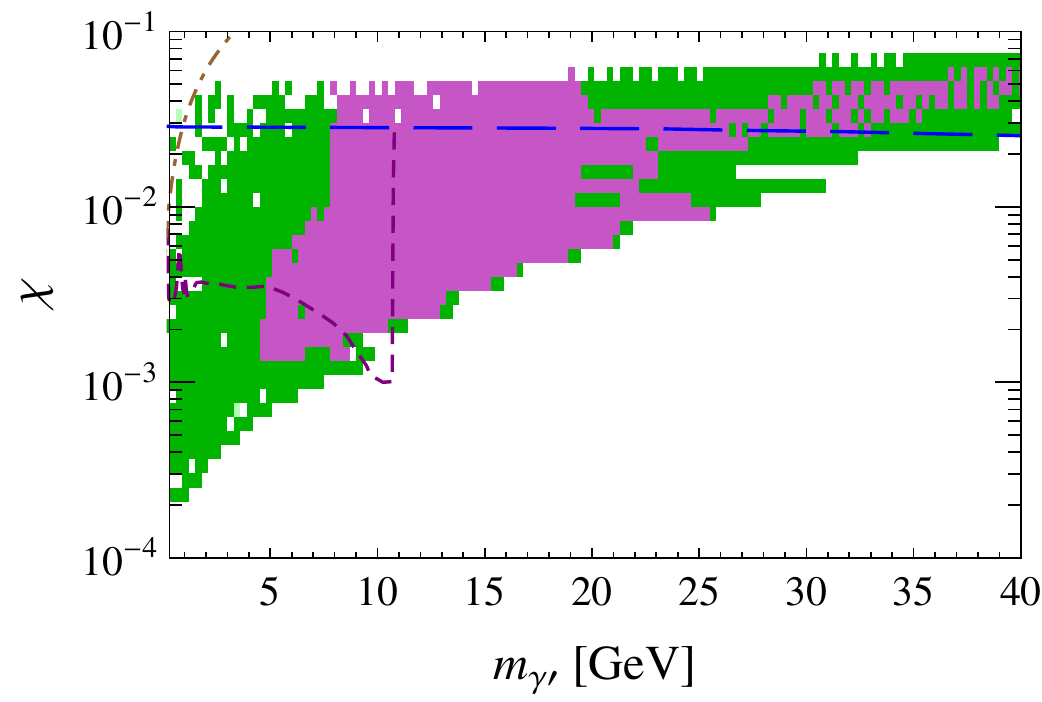}
\includegraphics[width=0.33\textwidth]{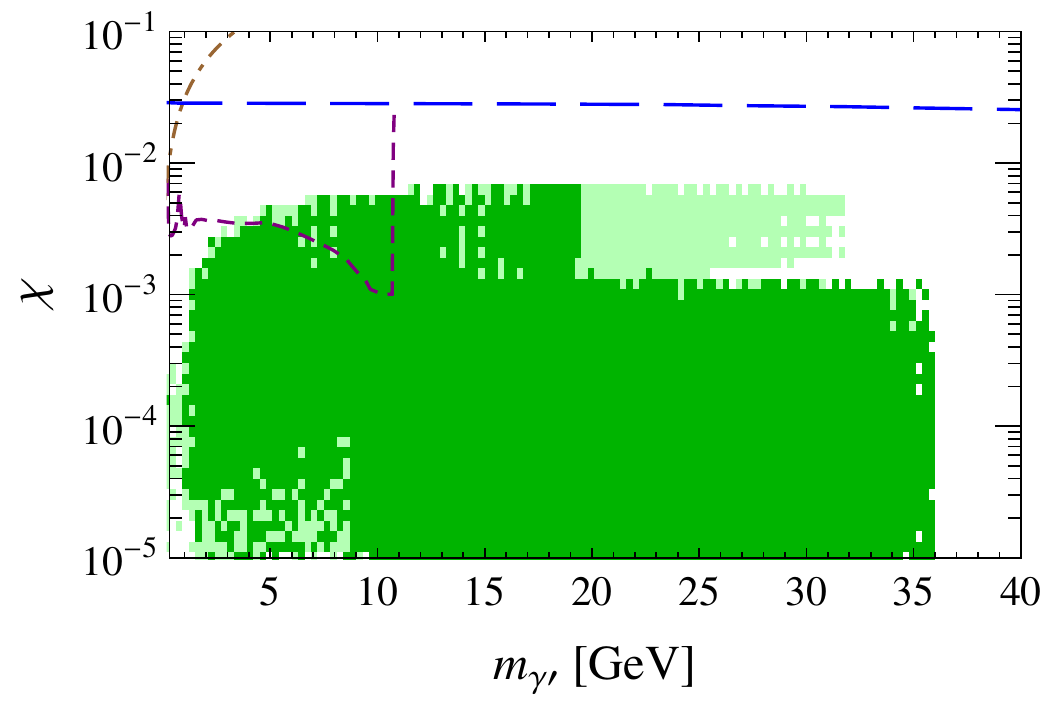}}
\vspace{-0.4cm} \caption{Scatter plot for the supersymmetric hidden sector with breaking induced by the visible sector \textit{(left)} or radiatively \textit{(right)}.} \label{fig-SUSYScattPlot}
\vspace{-0.35cm}
\end{wrapfigure}
the way the hidden gauge symmetry is broken. If the hidden gauge symmetry is broken via the effective Fayet-Iliopoulos term induced in the hidden sector through kinetic mixing with the visible Higgs D-term, the DM can be either a Dirac or a Majorana fermion. The former has SI scattering with a similar phenomenology as in the case of the toy-model. Scanning over a range of DM masses and $0.1\leq\kappa\leq10$, the scatter plot on the left of Fig.~\ref{fig-SUSYScattPlot} shows again in dark green the points that give the correct relic abundance and in purple those consistent with CoGeNT for a Standard Halo Model. The Majorana fermion on the other hand has mostly axial couplings and therefore the scattering is spin-dependent (SD) and only weakly constraint by direct detection experiments. Because of the Higgs-portal term (cf.~\cite{Foot:1991bp,Andreas:2011in}) SI scattering is possible but cross sections are several orders of magnitude below and without any chance of explaining signals in SI direct detection experiments . In the case where the hidden gauge symmetry breaking is induced by the running of the Yukawa coupling $\lambda_S$, a Majorana fermion remains as lightest particle in the spectrum. As shown in Fig.~\ref{fig-SUSYScattPlot}, it can produce the total relic abundance (dark green) or give a subdominant contribution (light green) but as the scattering is again mostly SD it is not able to account for the signals in SI direct detection experiments.

\section{Conclusions}
Hidden sectors are motivated both from a top-down (string theory, SUSY) and bottom-up (g-2, DM) perspective. Possibly containing gauge and matter fields they can be phenomenologically of interest for dark forces and DM. There is much experimental activity to constrain hidden photons. We presented here an overview of the latest limits. For hidden sectors containing also DM we showed that a simple toy-model with Dirac fermion DM is consistent with relic abundance and direct detection requirements and additionally able to have the correct cross section for CoGeNT. A more sophisticated supersymmetric hidden sector gives a similar DM phenomenology if the breaking of the hidden gauge symmetry is induced by the visible sector. Furthermore, in this case or for radiatively induced breaking, it is also possible to have a Majorana fermion with dominant SD scattering as DM candidate. Thus, supersymmetric models with gravity mediation give viable DM candidates with interesting signatures in experiments.

\subsubsection*{Acknowledgements}
\vspace{-0.1cm}
\noindent This work was done together with Mark Goodsell, Carsten Niebuhr and
Andreas Ringwald.


\begin{footnotesize}

\end{footnotesize}


\end{document}